\documentclass[prb,superscriptaddress,twocolumn,showpacs,amsmath,amssymb]{revtex4}
\usepackage{graphicx} 
\usepackage{bm}
\usepackage{txfonts}
\usepackage{mathrsfs}

\def\b0{{\mathbf 0}}

\def\b0{{\mathbf 0}}

\begin{document}

\title{Renormalization theory for the FFLO states at $T>0$}
\author{Pawel Jakubczyk}
\affiliation{Institute of Theoretical Physics, Faculty of Physics, University of Warsaw, Pasteura 5, 02-093 Warsaw, Poland}
\affiliation{Max-Planck-Institute for Solid State Research,
 Heisenbergstr.\ 1, D-70569 Stuttgart, Germany}
%
%
\date{\today}
\begin{abstract}
Within the renormalization group framework we study the stability of superfluid density wave states, known as Fulde-Ferrell-Larkin-Ovchinnikov (FFLO) phases, with respect to thermal order-parameter fluctuations in two and three-dimensional 
($d\in \{2,3\}$)
systems. 
We analyze the renormalization-group flow of the relevant ordering wave-vector $\vec{Q_0}$. The calculation indicates an instability of the FFLO-type states towards either a uniform superfluid or the normal state in $d\in\{2,3\}$ and $T>0$.  
In $d=2$ this is signaled by $\vec{Q_0}$ being renormalized towards zero, corresponding to the flow being attracted either to the usual Kosterlitz-Thouless fixed-point or to the normal phase. 
We supplement a solution of the RG flow equations by a simple scaling argument, supporting the generality of the result. 
The tendency to reduce the magnitude of $\vec{Q_0}$ by thermal fluctuations persists in $d=3$, where the very presence of long-range order is immune to thermal fluctuations, but the  effect of attracting $\vec{Q_0}$ towards zero  
by the flow remains observed at $T>0$.

\end{abstract}

\pacs{03.75 Ss, 67.85.-d, 74.20.Mn }

\maketitle


\section{Introduction}
The recent progress in manipulating cold atomic gases triggered the renaissance of interest in the Fulde-Ferrell-Larkin-Ovchinnikov (FFLO)\cite{Fulde64, Larkin64} phases. Such states may occur when a number (and/or mass) imbalance between the distinct fermion species 
forming the Cooper pairs is imposed. This gives rise to a mismatch between the Fermi surfaces of the two relevant particle species. 
For a sufficiently large imbalance pairing is suppressed and the superfluid phase becomes completely expelled from the phase diagram. The imbalance provides therefore an experimentally accessible non-thermal control 
parameter allowing for tuning the system across a quantum phase transition. A number of phases were proposed as candidates for intermediate stable states at the edge of the transition to the normal state. 
In the FFLO scenario (see e.g. Refs.~[\onlinecite{Fulde64, Larkin64, Combescot01, Bowers02, Casalbuoni04, Combescot04, Mora05, Mizushima05, He06, Sheehy07, Dey09, Ptok09, Yanase09, Baarsma10, Radzihovsky10, Ptok13, Cai11, Rosenberg15, Roscher15, Sheehy15, Piazza16, 
Toniolo17, Ptok17}]) pairing occurs at a finite momentum $\vec{Q_0}$ giving rise to a spatially nonuniform superfluid. Such a possibility was first discussed in the solid-state physics setup, where the imbalance is introduced by a magnetic 
field splitting the spin-up and spin-down bands. Potential realizations were later proposed in a diversity of contexts such as ultracold gases (e.g.~[\onlinecite{Radzihovsky10, Baarsma10, Cai11, Sheehy15, Roscher15, Toniolo17}]), high $T_c$ 
superconductors (e.g.~[\onlinecite{Yanase09}]), organic superconductors (e.g.~[\onlinecite{Piazza16}]), 
or quantum chromodynamics (e.g.[\onlinecite{Bowers02, Casalbuoni04}]).

Most of these studies relied on mean-field type treatments, and until recently considerably less attention was paid to the question of stability of the FFLO phases to thermal and quantum fluctuations. As was presumably for the first time noticed 
in Ref.~[\onlinecite{Shimahara98}] for the case of $d=2$, the FFLO states are fragile due to the presence of the Goldstone modes. The structure of the Goldstone modes' spectra in the FFLO states was somewhat later investigated in more detail in 
Ref.~[\onlinecite{Samokhin10}]. The remarkable contribution of Refs.~[\onlinecite{Radzihovsky09, Radzihovsky11}] predicted the instability of true long-range order in the Larkin-Ovchinnikov state in $d=3$ caused by an additional Goldstone mode related to 
translational symmetry-breaking. It also pointed out enlightening connections to the physics of classical liquid crystals. Ref.~[\onlinecite{Yin14}] addressed the stability of the Fulde-Ferrell state to Gaussian fluctuations in $d=2$, also pointing at 
a possible instability at $T>0$, while Ref.~[\onlinecite{Devreese14}] found in $d=3$ only small corrections to the region of stability of the FFLO states from phase fluctuations. A very recent study [\onlinecite{Wang17}] suggests a generic instability of the FFLO states at any $T>0$ due to pairing fluctuations both in $d=2$ and $d=3$. 

In this paper we address the problem of stability of the non-uniform FFLO-like superfluids to order-parameter fluctuations at $T>0$ applying a renormalization group framework. 
Using an effective order-parameter  action as a starting point, we follow the renormalization-group (RG) flow of the ordering wave-vector $\vec{Q_0}$. We find a tendency to reduce the magnitude of $\vec{Q_0}$ to zero by fluctuations. 
For two-dimensional systems we give a simple and rather general analytical argument for the instability of the FFLO-type states. A detailed analysis of the RG flow equations, which are derived under certain assumptions, 
leads to the conclusion that the effect of 
reducing $\vec{Q_0}$ to zero persists also in $d=3$. This is supported by a direct numerical solution of the analyzed RG flow equations. The emergent mechanism for destruction of the FFLO states is distinct from that discussed in 
Ref.~[\onlinecite{Radzihovsky11}]. In particular it does not invoke the specific anisotropic features of the Goldstone propagator. It also assumes the presence of only one (superfluid) Goldstone mode. We believe the conclusion of our study holds for a 
set of superfluid density wave states, which is broader than the Fulde-Ferrell and Larkin-Ovchinnikov classes. We however restrict our present study to the effects of thermal fluctuations, leading aside quantum effects. Our study 
therefore does not apply to $T=0$. Importantly, the 
paper is restricted to neutral superfluids, such as those realized in cold atomic gases. We make here no claim concerning charged systems, where the presence of the electromagnetic field leads to gapping the Goldstone mode and stabilizes 
the FFLO state. We make also the observation that the experimental evidence supporting the FFLO states seems to be available exclusively for charged systems, such as organic superconductors 
(see e.g. Refs.~[\onlinecite{Uji12, Uji13, Tsuchiya15, Koutroulakis16}]).  

The paper is organized as follows: In Sec.~II we introduce the relevant order parameter action. We review the mean-field theory and the effective action for Goldstone fluctuations. The cases of Fulde-Ferrell (FF) and Larkin-Ovchinnikov (LO) states 
lead to 
fluctuation spectra of distinct nature and require separate treatments. In Sec.~III we introduce a simplified model encompassing the important features of both the FF and LO states, but characterized by weaker fluctuation properties. In the subsequent 
Sec.~IV we present the RG framework employed to tackle the problem of the renormalization of the ordering wave vector due to order-parameter fluctuations. Sec.~V contains the technical details of the derivation of the RG flow equations, while in Sec.~VI we 
present their numerical solution in $d\in\{2,3\}$. We give our final conclusion in Sec.~VII.

\section{Effective action}
As a starting point to address the FFLO states at $\beta^{-1}=T>0$ we consider the scalar complex order parameter field $\phi(\vec{x})=\Delta(\vec{x}) e^{i\theta(\vec{x})}$ and the corresponding Ginzburg-Landau action: 
\begin{equation}
\label{action}
S[\phi] = \beta\int_{\vec{x}} \left[U(|\phi|^2) +\frac{1}{2}Z |\nabla \phi|^2 +\frac{1}{2}Y|\nabla^2\phi|^2+\frac{1}{2}X | \phi|^4(\nabla\theta)^2 \right].
\end{equation}
This form retains the effective potential $U(|\phi|^2)$ together with the relevant gradient terms. 
 The coefficient $Z$ is assumed negative, enforcing instability of the uniform order-parameter configuration towards a density-wave like phase. The system is stabilized by the term $\sim|\nabla^2\phi|^2$ with $Y>0$. 
 The last term allows for distinguishing between the phase and amplitude stiffness and may be given the interpretation in terms of a current-current interaction.\cite{Radzihovsky11} 
The system is $d$-dimensional, i.e. 
$\int_{\vec{x}}=\int_{V}d^dx$ and we consider the thermodynamic limit $V\to \infty$.
 Note that the action involves exclusively thermal fluctuations, restricting the analysis to $T>0$. In what follows we will parametrize the effective potential with the quartic form
\begin{equation}
U(|\phi|^2|) = m^2 |\phi|^2 +u |\phi|^4
\end{equation}
with $m^2<0$ and $u>0$. Truncating higher-order terms is not expected to influence our major conclusions. 
 A connection between the action of Eq.~(\ref{action}) and microscopic models may be made along the recognized track (see e.g. Refs.~[\onlinecite{Sheehy07, Baarsma10, Radzihovsky11}]). We refer to the previous works for a detailed discussion.
 
 We proceed by analyzing the two families of candidate ground states known as the Fulde-Ferrell and Larkin-Ovchinnikov phases. We note, however, that this list is far from complete and more complex superfluid wave states 
 (e.g. involving superpositions of non-collinear waves\cite{Combescot04}) may well be more stable at mean field (MF) level. The RG theory developed, under certain assumptions, in the subsequent sections is not restricted to the FF and LO classes. 

 \subsection{FF states}
The Fulde-Ferrell ansatz proposes the superfluid order parameter in the plane-wave form 
\begin{equation}
\phi_{FF}(\vec{x})=\Delta(\vec{x})e^{i\vec{Q}\vec{x}} e^{i\theta(\vec{x})}\;.
\end{equation}
In the MF picture the amplitude $\Delta$ and phase $\theta$ are assumed constant. Plugging $\phi_{FF}(\vec{x})$ into Eq.~(\ref{action}) one obtains
\begin{equation}
S^{MF}_{FF}(\Delta, \vec{Q})= \beta\int_{\vec{x}} \left[\left(m^2+\frac{1}{2}Z\vec{Q}^2 +\frac{1}{2}Y\vec{Q}^4\right)\Delta^2 + (u+\frac{1}{2}X\vec{Q}^2)\Delta^4\right]  \;.  
\end{equation} 
The ordering wavevector $\vec{Q}=\vec{Q_0}$ may now be related to the system parameters by extracting the minimum of $S^{MF}_{FF}(\Delta, \vec{Q})$ with respect to $\vec{Q}$. This yields
\begin{equation} 
\label{Q_0}
\vec{Q_0}^2=-\frac{Z+X\Delta^2}{2Y} \Theta[-(Z+X\Delta^2)]\;,
\end{equation} 
where $\Theta$ denotes the Heaviside function. The MF order parameter $\Delta_0$ follows from the minimum of $S^{MF}_{FF}(\Delta, \vec{Q}=\vec{Q_0})$ with respect to $\Delta$. We restrict here to situations where the system is ordered 
($\Delta_0^2>0$) at mean-field level. 

 We proceed by analyzing the effective action for phase fluctuations around the MF introduced above
\begin{equation}
S^{Fl}_{FF}[\theta]=S\left[\phi(\vec{x})=\Delta_0 e^{i\vec{Q_0}\vec{x}}e^{i\theta(\vec{x})}\right]\;.
\end{equation}
The amplitude remains fixed, but the phase degree of freedom $\theta$ (corresponding to the Goldstone mode) is allowed to fluctuate. Working out the derivatives,   
 we obtain   
\begin{equation} 
\begin{aligned}
\label{action_FF}
S^{Fl}_{FF}[\theta]={}  \beta\int_{\vec{x}}  \Bigg[ & \left(m^2+\frac{1}{2}\vec{Q_0}^2\left(Z-\frac{1}{2}(Z+X\Delta_0^2) \right)\right)\Delta_0^2+ \\
                                    &       \left(u+\frac{1}{2}X\vec{Q_0}^2\right)\Delta_0^4 \\
                                   & - (Z+X\Delta_0^2)\Delta_0^2\left(\partial_{x_1}\theta\right)^2 + \frac{1}{2}Y\Delta_0^2\left(\nabla^2\theta\right)^2 + \mathcal{R}  \Bigg]\;,
\end{aligned}
\end{equation}
where we chose 
\begin{equation}
\vec{Q_0}=\sqrt{-\frac{Z+X\Delta_0^2}{2Y}}\vec{e}_{x_1} 
\label{Q_0_2}
\end{equation} 
 and the remainder $\mathcal{R}$ contains higher-order derivative terms.  The first two terms are merely constants but we keep them here for the sake of the argument given below.
As discussed in Refs.~[\onlinecite{Shimahara98, Radzihovsky09, Radzihovsky11}], 
vanishing of the phase stiffness $\rho_{0, \perp}$  
in the directions perpendicular to $\vec{Q_0}$ is a peculiarity of the FF state and clearly amplifies the role played by fluctuations. 

The full RG analysis will be carried out in Sec.~IV-VI. For $d=2$ one may however infer important information from the very structure of Eq.~({\ref{action_FF}) invoking known statistical physics facts. 
Let us therefore now examine the structure of Eq.~(\ref{action_FF}) in $d=2$ from the point of view of scaling and renormalization theory.  
At high RG scales ($k$) the coefficient 
\begin{equation}
m_0^2\equiv   m^2+\frac{1}{2}\vec{Q_0}^2\left(Z-\frac{1}{2}(Z+X\Delta_0^2) \right)
\end{equation}
of $\Delta_0^2$ is negative, but increases towards zero when the scale is lowered. This is because fluctuation effects generically tend to reduce ordering tendencies. If $m_0^2$ crosses zero at 
a finite RG scale, the system goes into the normal (i.e. nonsuperfluid) phase. The other possibility is that ${m_0}^2$ approaches zero from below asymptotically. The scenario where $m_0^2$ neither crosses nor approaches zero would imply the presence 
of long-range order and therefore contradict the Mermin-Wagner 
theorem.\cite{Mermin66} The asymptotic vanishing of $m_0^2$ for $k\to 0$ does not yet imply the independent vanishing of each of the two terms composing it. However, there is no obvious reason for any cancellation in the flow of these two quantities and 
it is most natural to assume that each of them vanishes independently during the RG flow. This is also confirmed in numerical solutions of the flow equations (see Sec.~VI). We therefore write 
\begin{equation} 
\label{purport}
 \lim_{k\to 0} \left[\vec{Q_0}^2\left(Z-\frac{1}{2}(Z+X\Delta_0^2) \right)\right]=0\;.
\end{equation}
Now observe that by Eq.~(\ref{action_FF}) the quantity $\rho_{0,\parallel}\equiv-(Z+X\Delta_0^2)\Delta_0^2$ is the phase stiffness along $\vec{Q_0}$, which under RG flow must approach a constant to support a superfluid phase in $d=2$. Since (due to Mermin-Wagner theorem) $\Delta_0\to 0$ as $k\to 0$, 
one has $-(Z+X\Delta_0^2)\to\infty$. 
Assuming no unexpected cancellations in Eq.~(\ref{purport}), this suffices for writing 
\begin{equation}
\lim_{k\to 0} \vec{Q_0}^2 = 0\,
\end{equation}    
which is the way of avoiding violation of Eq.~(\ref{purport}) and implies the unavoidable renormalization of the ordering wavevector $\vec{Q_0}$ to zero. 
The argument does not use the 'anisotropic' feature of Eq.~(\ref{action_FF}) inherited from the specific character of the FF state and holds equally well for other cases such as the LO state discussed below. 
It is however crucial to restrict to $d=2$ since otherwise $\Delta_0^2$ remains positive, $-(Z+X\Delta_0^2)$ does not diverge in the low-$T$ phase and, in consequence, the above argument gets spoiled.  In fact, as we see in the RG theory, the physics underlying the above argument is governed 
by the anomalous dimension $\eta$, since $\Delta_0^2\sim k^\eta$ and therefore $(Z+X\Delta_0^2)\sim k^{-\eta}$ alike in the standard Kosterlitz-Thouless theory. The other important point to note for future reference is that, by Eq.~(\ref{Q_0_2}), the vanishing of $\vec{Q_0}$ under the RG flow is possible only if $(Z+X\Delta_0^2)\to 0$ or $Y\to\infty$ as $k\to 0$. Observe also that $\vec{Q_0}^2$ must vanish faster than $k^\eta$ as a function of $k$.

\subsection{LO states}
The Larkin-Ovchinnikov ansatz restricts the order-parameter configurations to standing waves 
\begin{equation}
\phi_{LO}(\vec{x})=\Delta(\vec{x})\cos\left(\vec{Q}\vec{x}+\gamma(\vec{x})\right) e^{i\theta(\vec{x})}\;.
\end{equation}
In comparison to the FF case one anticipates here the presence of an extra Goldstone mode $(\gamma)$ related to spontaneous breaking of translational symmetry. 
The MF analysis parallels the FF case. Assuming $\Delta(\vec{x})$, $\gamma(\vec{x})$, $\theta(\vec{x})$ to be constant, plugging $\phi_{LO}$ into Eq.~(\ref{action}), and minimizing the result yields 
\begin{equation}
\vec{Q_0}^2=-\frac{Z}{2Y}\Theta[-Z]\;. 
\end{equation}
There is however a crucial difference in the structure of the effective action for the Goldstone modes. By assuming a frozen amplitude $\Delta(\vec{x})=\Delta_0$ we plug $\phi_{LO}$ into Eq.~(\ref{action}). Executing the derivatives and dropping 
subleading oscillatory terms we obtain
\begin{equation}
\begin{aligned}
 S^{Fl}_{LO}[\theta, \gamma]={} \beta \int_{\vec{x}}\Bigg[ & \tilde{U}(\Delta_0^2) +  \rho_{0,\parallel}(\partial_{x_1}\theta)^2+\rho_{0,\perp}(\partial_{\perp}\theta)^2 \\
                                                                             & -\frac{1}{2}Z\Delta_0^2 (\partial_{x_1}\gamma)^2+\frac{1}{4}Y\Delta_0^2(\nabla^2\gamma)^2 + \mathcal{R} \Bigg]\;.
\end{aligned}
\end{equation}
Here $\tilde{U}(\Delta_0^2)$ is a constant. The superfluid stiffness $\rho_{0,\parallel}=f(Z,X)\Delta_0^2$ along $\vec{Q_0}$ is controlled by both $Z$ and $X$, while the perpendicular phase stiffness $\rho_{0,\perp}\sim X\Delta_0^4$ is fully 
determined by the $X$-term. The precise form of the function $f(Z,X)$ is not important for us now. Notably, the fluctuations of the $\gamma$ mode are controlled by terms quartic in derivatives in the direction orthogonal to $\vec{Q_0}$. The implications of this fact for the LO state are discussed in detail in 
Ref.~[\onlinecite{Radzihovsky11}].  
 
Clearly, the fluctuation properties of the competing superfluid density wave phases (such as the FF and LO states) are sensitive to the particular forms of these ground states. The LO state hosts two Goldstone modes, while the FF state is 
characterized by a vanishing stiffness $\rho_{0,\perp}$ in the perpendicular directions. In the following sections we will show, however that the superfluid order parameter fluctuations may lead to an instability of the FFLO-type states
even in the absence of such special features (i.e for models with one Goldstone mode characterized by quadratic spectrum in all directions) at any finite temperature $T>0$ in $d\in \{2,3\}$. This is in line with the argument presented in 
Sec.~IIA.
 
\section{Model} 
We now introduce a simplified effective model that encompasses features of both the FF and LO states, but is characterized by weaker fluctuation effects compared to those occurring in the models discussed above. We describe in detail its relation 
to the FF model from the previous section, and then also discuss its connection to the LO state.  

Consider first the FF ansatz with both amplitude and phase fluctuations 
\begin{equation}
 \phi(\vec{x}) =\left(\alpha +\sigma(\vec{x})\right)e^{i\vec{Q}\vec{x}}e^{i\theta(\vec{x})}\;.
\end{equation}
Plugging this into Eq.~(\ref{action}) and proceeding along the line of Sec.~II yields the effective action for the $\sigma$ and $\theta$ modes: 
\begin{equation}
\begin{aligned}
S[\sigma,\theta] = {} \beta\int_{\vec{x}}\Bigg\{ & \left[m^2+\frac{1}{2}Z \vec{Q_0}^2+\frac{1}{2}Y\vec{Q_0}^4\right](\alpha+\sigma)^2  + \\ 
                                            &\left[u+\frac{1}{2}X\vec{Q_0}^2\right](\alpha+\sigma)^4 + \\
                                            &\tilde{Z}(\alpha+\sigma)^2(\partial_{x_1}\theta)^2 + \tilde{Z}(\partial_{x_1}\sigma)^2 + \\
                                            &\frac{1}{2}Z(\nabla_{\perp}\sigma)^2+\frac{1}{2}Y\left[(\alpha+\sigma)^2(\nabla^2\theta)^2+(\nabla^2\sigma)^2\right] \Bigg\}\;,
\end{aligned}
\end{equation}
where $\vec{Q_0}^2=-\frac{Z+X(\alpha+\sigma)^2}{2 Y}\approx -\frac{Z+X\alpha^2}{2 Y}$ is assumed positive and $\tilde{Z}=-[Z+X(\alpha+\sigma)^2]\approx -[Z+X\alpha^2]>0$. We now trade the $\theta$ mode for the transverse mode ($\pi$) 
by writing 
\begin{equation}
\phi (\vec{x})\approx [\alpha +\sigma(\vec{x})+i\alpha\theta(\vec{x})]e^{i\vec{Q}\vec{x}}=[\alpha+\sigma(\vec{x})+i\pi(x)]e^{i\vec{Q}\vec{x}}
\end{equation}
with $\pi(\vec{x})=\alpha \theta(\vec{x})$. We now additionally simplify the model by neglecting the 
anisotropy of the amplitude stiffness and {\it reduce} the phase fluctuations by assuming a finite stiffness $\tilde{Z}$ both along and perpendicular to $\vec{Q_0}$. Recalling that $(\alpha+\sigma)^2=|\phi|^2$, we introduce $\rho\equiv \frac{1}{2}|\phi|^2$ 
and reparametrize the uniform (Mexican-hat shaped) uniform part of the action. This leads us to the following form
\begin{equation}
\begin{aligned}
\label{S}
  S_{eff}[\sigma, \pi] ={} \beta \int_{\vec{x}} \Bigg\{ & \frac{\lambda}{2}(\rho-\rho_0)^2   +\frac{1}{2}Z_\pi (\nabla \pi)^2 + \frac{1}{2}Z_\sigma (\nabla \sigma)^2 \\
                        &+ \frac{1}{2}Y\left[(\nabla^2 \pi)^2+(\nabla^2 \sigma)^2\right] \Bigg\}\;,
\end{aligned}
 \end{equation}
which is parametrized by the interaction coupling $\lambda$, the effective potential minimum $\rho_0=\frac{1}{2}\alpha^2$ and the three (positive) gradient coefficients. In the present calculation 
we have $Z_\sigma=Z_\pi=2\tilde{Z}$. Crucially, the ordering wavevector is given by 
\begin{equation}
\label{Q_0}
\vec{Q_0}^2 = \frac{\tilde{Z}}{2Y}=\frac{Z_\pi}{Y} \;.
\end{equation}
The above shows that Eq.~(\ref{S}) may be considered as a variation of the FF action from the previous section, where the Goldstone mode becomes weakened by assuming a finite stiffness $\rho_{0,\perp}$ in the direction orthogonal to 
$\vec{Q_0}$. The presence of the massive amplitude mode $\sigma$ has a minor effect on our further results. 

We additionally observe that Eq.~(\ref{S}) may also be interpreted as a modification of the LO action, where the translational Goldstone mode $\gamma$ is disregarded together with the anisotropy of the superfluid stiffness coefficients. The relation 
Eq.~(\ref{Q_0}) is valid also for the LO system. One may reliably assume that fluctuation effects are again reduced as compared to $S^{Fl}_{LO}$ since we take out one massless mode. The structure of the action given by Eq.~(\ref{S}) is typical for  
order-parameter theories with a $U(1)$ symmetry, which are well-studied in many contexts. The present problem requires however an analysis of the RG flow of the Laplacian coefficient $Y$ in order to access the impact of the order parameter fluctuations 
on the ordering wavevector $\vec{Q_0}$ given by Eq.~(\ref{Q_0}). 

\section{RG theory} 
The action defined in Eq.~(\ref{S}) is the starting point of the RG analysis. We derive the relevant flow equations for the set of couplings $\{\rho_0, \lambda, Z_\pi, Y\}$ from the functional RG in the one particle irreducible 
formulation.\cite{Wetterich93} The flow of the ordering wavevector $\vec{Q_0}$ is then extracted from Eq.~(\ref{Q_0}). As already remarked, we put $Z_\sigma=Z_\pi$.  The central object of the formalism is the scale ($k$) dependent effective action $\Gamma_k[\phi]$ which interpolates 
between the bare action Eq.~(\ref{S}) at high momentum scales ($k=k_0$) and the full free energy for $k\to 0$. The functional flow upon varying the scale $k$ is governed by the Wetterich equation\cite{Wetterich93}
\begin{equation}
\label{Wetterich_eq}
\partial_k \Gamma_k[\phi] = \frac{1}{2}{\rm Tr} \left\{ \partial_k R_k(q) \left[\Gamma_k^{(2)}[\phi]+R_k(q)\right]^{-1}\right\}\;.
\end{equation}
Here the trace (Tr) sums over momenta $(\vec{q})$ and the field components $(\{\sigma, \pi\})$, $\Gamma_k^{(2)}[\phi]$ is the second functional derivative of $\Gamma_k[\phi]$, and the cutoff function $R_k(q)$ is an artificial mass term added to the 
inverse propagators (both in the $\sigma$ and $\pi$ directions) to suppress the modes below the scale $k$. It adds a mass $\sim k^2$ to modes with $ q \equiv |\vec{q}|\ll k$, but leaves the modes with $q\gg k$ unaffected.
The quantity $\Gamma_k[\phi]$ may be understood as the free energy including fluctuation modes between $k$ and the highest cutoff scale $k_0$. For $k=k_0$ all 
fluctuations are frozen $(R_{k\to k_0}(q)\to\infty)$ and in consequence $\Gamma_k[\phi]\to S_{eff}[\phi]$. In the opposite limit we have $R_{k\to 0}(q)\to 0$ and all fluctuations are included into the partition function. The Wetterich framework was 
successfully used in a diversity of contexts over the last years (for reviews see e.g.~[\onlinecite{Berges02, Pawlowski07, Kopietz10, Delamotte12, Metzner12}]). The imbalanced Fermi gases were addressed with this approach in 
Refs.~[\onlinecite{Roscher15, Floerchinger10, Strack14, Boettcher15_1, Boettcher15_2}].

In what follows we promote the set of couplings $\{\rho_0, \lambda, Z_\pi, Y\}$ to functions $\{\rho_{0,k}, \lambda_k, Z_{\pi,k}, Y_k \}$ of the cutoff scale $k$ and parametrize the flowing action $\Gamma_k[\phi]$ with the form given by Eq.~(\ref{S}). By plugging this parametrization 
into Eq.~(\ref{Wetterich_eq}), we project the functional flow onto a finite set of flow equations describing the evolution of the couplings. This modest approximation captures the effect of the wavevector renormalization, which is at 
present 
of 
our main interest. Note that the calculation requires accounting for terms that are 4th order in the derivative expansion\cite{Canet03} ($\sim q^4$), but does not treat contributions $\sim q^4$ which are not captured by the 2-point function.  
In addition there is an important assumption underlying the calculation, that the two-point function may be parametrized by the quartic form for all the scales $k$ and within the whole relevant domain of momenta $q$. Observe, that the argument given in Sec.~IIA relies on this parametrization as well. For the imbalanced 
Fermi gases at $T>0$ this appears plausible in view of the MF results.\cite{Baarsma10} However, the present calculation may not be understood as a first step in a systematic procedure of expanding the two-point function in momenta around the flowing
 ordering wavevector. Indeed, going to higher-orders in gradients would lead to increasingly inaccurate parametrization of the two-point function except for the immediate vicinity of $Q_0$. This is an important point of the 
analysis, since we are dealing with nonuniversal properties governed to a large extent (within the present truncation) by the irrelevant coupling $Y$. 
 We expect that the present approximation may break down if $\vec{Q_0}$ is located sufficiently far from zero. Within a more accurate treatment one might employ the so-called BMW scheme\cite{Blaizot06, Benitez12, Rose15} of functional RG. In that 
 framework the flow of the full momentum dependence of the two-point function is computed without a recourse to any expansions. We believe this is the most appropriate way to address the problem (in $d=3$ in particular). We notice however, that the BMW procedure suffers substantial numerical difficulties deep in the low-$T$ phases.  
 
The technical details of the projection of the flow are described in Sec.~V, which is followed by a numerical integration of the resulting flow equations in Sec.~VI. 

\section{RG flow equations}
Here we describe the procedure to derive the flow equations, first focusing on the uniform part characterized by $\{\rho_{0,k}, \lambda_k\}$ and subsequently analyzing the gradient parameters $\{Z_{\pi,k}, Y_k\}$. Our parametrization 
implies that the $\sigma$ and $\pi$ inverse propagators take the following form
\begin{equation}
G_\sigma^{-1}(q) \equiv \Gamma^{(2)}_{\sigma \sigma}(q) = m_k^2+Z_{\sigma,k}q^2+Y_k q^4+R_k(q)\;,
\end{equation} 
and
\begin{equation}
G_\pi^{-1}(q) \equiv \Gamma^{(2)}_{\pi \pi}(q) = Z_{\pi,k}q^2 + Y_k q^4 +  R_k(q)\;.
\end{equation}
The longitudinal mass $m_k^2$ is given by $m_k^2=2\rho_{0,k}\lambda_k$ and $q=|\vec{q}|$.
\subsection{Effective potential flow}
The flow of the effective potential is extracted by the standard procedure\cite{Berges02} of evaluating Eq.~(\ref{Wetterich_eq}) on a uniform field configuration, $\rho={\rm const}>0$. This reads 
\begin{equation}
\label{U_flow}
 \partial_k U_k(\rho) = \frac{1}{2}\int_{\vec{q}} \partial_k R_k(q)G_\pi(q,\rho) + \frac{1}{2}\int_{\vec{q}} \partial_k R_k(q)G_\sigma(q,\rho)\;,
\end{equation}
where 
\begin{equation}
G_{\sigma}^{-1}(q,\rho)= U_k'(\rho)+2\rho U_k''(\rho)+ Z_{\sigma,k}q^2+Y_k q^4+R_k(q)\;,
\end{equation}
\begin{equation}
G_{\pi}^{-1}(q,\rho)= U_k'(\rho) + Z_{\pi,k}q^2+Y_k q^4+R_k(q)\;,
\end{equation}
and 
\begin{equation}
 \int_{\vec{q}} = T\int\frac{d^d q}{(2\pi) ^d}\;.
\end{equation}
We extract the flow of the quartic coupling $\lambda_k$ by differentiating Eq.~(\ref{U_flow}) twice with respect to $\rho$ and evaluating at $\rho=\rho_0$. The observation that $\lambda_k = U_k''(\rho=\rho_0)$ leads to 
\begin{equation}
\partial_k \lambda_k =\lambda_k^2 \int_{\vec{q}} \partial_k R_k(q)G_\pi^3(q) + 9\lambda_k^2 \int_{\vec{q}} \partial_k R_k(q)G_\sigma^3(q)\;. 
\label{lambda_flow}
\end{equation}
The flow of $\rho_{0,k}$ is obtained from 
\begin{equation}
 0=\frac{d}{dk}U_k'(\rho)|_{\rho=\rho_0}=\frac{\partial U_k'(\rho)}{\partial k}|_{\rho=\rho_0}+  \frac{\partial U_k'(\rho)}{\partial \rho}|_{\rho=\rho_0}\frac{d \rho_0}{dk}\;,
\end{equation}
where the term $\frac{\partial U_k'(\rho)}{\partial \rho}|_{\rho=\rho_0}$ follows from executing the derivative of Eq.~(\ref{U_flow}) with respect to $\rho$ and evaluating at $\rho=\rho_0$. This gives  
\begin{equation}
 \partial_k \rho_{0,k} =\frac{1}{2} \int_{\vec{q}} \partial_k R_k(q)G_\pi^2(q) + \frac{3}{2} \int_{\vec{q}} \partial_k R_k(q)G_\sigma^2(q)\;.
 \label{rho_0_flow}
\end{equation}
Eq.~(\ref{lambda_flow}) and Eq.~(\ref{rho_0_flow}) describe the flow of the effective potential within the present truncation and need to be supplemented by the flow equations for the gradient coefficients, which are derived below. 
\subsection{Propagator flow}
The flow equation for the inverse propagator is derived by taking the second functional derivative of Eq.~(\ref{Wetterich_eq}), which is then evaluated at $\phi=const>0$. 
The first step leads to 
\begin{equation}
\label{prop_flow}
\partial_k \Gamma^{(2)}_{\phi^j_{q_2}\phi^i_{q_1}} = {\rm Tr} \left\{-\mathbb{S}   \Gamma^{(3)}_{\phi^j_{q_2}\phi\phi}\left(\Gamma^{(2)}_{\phi\phi}\right)^{-1} \Gamma^{(3)}_{\phi^i_{q_1}\phi\phi} + 
\frac{1}{2} \mathbb{S}\Gamma^{(4)}_{\phi^j_{q_2}\phi^i_{q_1}\phi\phi}   \right\}\;,
\end{equation}
with the single-scale propagator $\mathbb{S}\equiv -\left(\Gamma^{(2)}\right)^{-1}\partial_k R_k \left(\Gamma^{(2)}\right)^{-1}$. We dropped the scale and internal momentum dependencies for clarity and $\phi\in\{\sigma, \pi\}$ is 
summed over. In addition $i,j\in\{1,2\}$, and $\phi^1=\sigma$, $\phi^2=\pi$.  Eq.~(\ref{prop_flow}) has a clear diagrammatic interpretation. We extract the flow of the $\pi$ mode propagator by specifying to $i=j=2$, evaluating at a constant field configuration, and choosing $\rho=\rho_0$. 
This yields 
\begin{equation}
\begin{aligned}
\label{Gamma_pi_flow}
 &\partial_k \Gamma^{(2)}_{\pi\pi}(q_1)= {}  \\
 & 2\rho_{0,k}\lambda_k^2 \int_{\vec{q}}\partial_k R_k(q)\left[G_\sigma^2(q)G_\pi(|\vec{q}+\vec{q_1}|)+G_\pi^2(q)G_\sigma(|\vec{q}+\vec{q_1}|)\right]\;.
\end{aligned}
\end{equation}
We now expand $G_\sigma(|\vec{q}+\vec{q_1}|)$ and $G_\pi(|\vec{q}+\vec{q_1}|)$ around $\vec{q_1}=0$, retaining terms up to order $\sim\vec{q_1}^4$ and subsequently perform the angular integrations on the right-hand side of Eq.~(\ref{Gamma_pi_flow}). 
This calculation yields the flow equations for 
$Z_{\pi,k}$ and $Y_k$, which are extracted as the $\vec{q_1}^2$ and $(\vec{q_1}^2)^2$ coefficients of the resulting expression. The obtained flow equations read: 
\begin{equation}
\begin{aligned}
\label{Z_flow}
& \partial_k Z_\pi = 2\rho_0\lambda^2 T \int_0^\infty dq (\partial_k R) G_\sigma^2 G_\pi^2 \Big\{\Big[-8Y \\ 
& +4(Z_\pi+2Yq^2)^2\left(G_\pi+G_\sigma\right)\Big]C_d q^{d+1} - \left(Z_\pi+2Yq^2\right)\frac{1}{\pi^{d-1}}q^{d-1}\Big\} ,
\end{aligned}
\end{equation}
and
\begin{equation}
\begin{aligned}
\label{Y_flow}
& \partial_k Y = {} 2\rho_0\lambda^2 T \int_0^\infty dq (\partial_k R) G_\sigma^2 G_\pi^2 \Bigg\{\\ 
&\Big[-2Y 
 +\left(Z_\pi+2Yq^2\right)^2\left(G_\pi+G_\sigma\right)\Big]\frac{1}{2\pi^{d-1}}q^{d-1} + \\
&\Big[\left(G_\pi+G_\sigma\right) 24Y \left(Z_\pi+2Yq^2\right) 
- \left(G_\pi^2+G_\sigma^2\right) 12 \left(Z_\pi+2Yq^2\right)^3\Big] C_dq^{d+1} \\
&+ \Big[\left(G_\pi+G_\sigma\right)16 Y^2 - \left(G_\pi^2+G_\sigma^2\right)12\left(Z_\pi+2Yq^2\right)^2(4Y) \\
&+ \left(G_\pi^3+G_\sigma^3\right)16 \left(Z_\pi+2Yq^2\right)^4 \Big]D_d q^{d+3}         \Bigg\}\;.
\end{aligned}
\end{equation}
To gain some notational clarity we dropped the scale and momentum dependencies and introduced $C_d=\frac{1}{2d\pi^{d-1}}$, $D_2=\frac{3}{16\pi}$, $D_3=\frac{1}{10\pi^2}$, originating from $\int_{\vec{q}}f(q)(\vec{q}\vec{q_1})^2=TC_d\int_0^\infty dq q^{d+1}f(q) \vec{q_1}^2$ 
and $\int_{\vec{q}}f(q)(\vec{q}\vec{q_1})^4=TD_d\int_0^\infty dq q^{d+3}f(q) (\vec{q_1}^2)^2$\;. The above equations together with Eq.~(\ref{lambda_flow}) and Eq.~(\ref{rho_0_flow}) form the set of coupled flow equations to be studied numerically in 
the subsequent section. We may however already at this point observe that the right-hand side of Eq.~(\ref{Y_flow}) contains terms with the momentum power lower by 2 as compared to the corresponding terms in Eq.~(\ref{Z_flow}). 
One then  naively anticipates $\partial_k Y_k\sim (\partial_kZ_{\pi,k})k^{-2}$, leading to $\vec{Q_0}^2\to 0$ for $k\to 0$ by virtue of Eq.~(\ref{Q_0}) under the condition that the flowing anomalous dimension $\eta\sim k\partial_kZ_{\pi,k}$ does not vanish too fast for $k\to 0$. This is always the case in $d=2$, since $\eta>0$ in the infrared limit $k\to 0$ in the entire low-temperature phase. Note that the picture is in perfect line with the argument from Sec.~IIA. As turns out, the above condition remains fulfilled also for $d=3$.
From this point of view the numerical results of the next section may appear expected. 
\subsection{Remarks on the functional RG truncation}
The projection of the Wetterich equation [Eq.~(\ref{Wetterich_eq})] described above goes along the path of earlier works and it is worthwhile discussing the anticipated accuracy of the results together with potential artefacts of the approximation. In $d=3$ the situation is rather simple [see e.g. Ref.~(\onlinecite{Berges02})] and one recovers the picture displaying a symmetry-broken phase separated from the normal phase by a phase transition. The properties of the transition itself (for example the values of the critical exponents) are captured at a qualitative level. Recovering numerically exact values requires a more sophisticated approximation level. We stress however that the present calculation does not require knowledge concerning the transition itself, since we are interested almost exclusively  in the properties of the low-$T$ phase. The situation in $d=2$ is somewhat more subtle, since the low-$T$ phase of relevance is the algebraic Kosterlitz-Thouless phase.\cite{Berezinskii71, Kosterlitz73} This may be relatively easily  obtained from the Gaussian theory of phase fluctuations.\cite{Chaikin_book} Is is somewhat striking that rather sophisticated functional RG truncations find it hard to exactly recover this result. This issue is discussed in detail in Ref.~[\onlinecite{Jakubczyk17}]. At the present truncation level, the line of RG fixed points characterizing the Kosterlitz-Thouless phase is recovered approximately (see the next section), such that the phase stiffness and anomalous dimension exhibit only very slow, logarithmic flow. The quality of the approximation improves when the temperature is lowered. As concerns the transition itself, the presently applied truncation level\cite{Graeter95} may serve to give an estimate of the transition temperature and the anomalous dimension. More sophisticated functional RG calculations, retaining terms up to infinite order in the order parameter field\cite{Gersdorff01, Jakubczyk14} reproduce accurately the universal properties (such as the phase stiffness jump, the anomalous dimension or the essential scaling of the correlation length in the high-$T$ phase), and even nonuniversal thermodynamics of specific microscopic models.\cite{Jakubczyk16} We stress again that the properties of the Kosterlitz-Thouless transition are not of major focus for us now, since we are most interested in the region of the phase diagram occupied by the low-$T$ phase. 
\section{Numerical integration of the RG flow}
A practical numerical solution of the flow equations requires specifying the cutoff function $R_k(q)$. Here we make the choice\cite{Litim01} 
\begin{equation}
\begin{aligned}
& R_k(q) = {} \\ 
&\left[Z_{\pi,k}\left(k^2-q^2\right)+Y_k\left(k^4-q^4\right)\right] \Theta \left[Z_{\pi,k}\left(k^2-q^2\right)+Y_k\left(k^4-q^4\right)\right]\;,
\end{aligned}
\end{equation}
which displays the properties required by the formalism, and, in addition allows for executing the integrals in the flow equations analytically. We introduce the logarithmic scale $s=-\log(k/k_{UV})$. Of our major interest are 
the parameters corresponding to the thermodynamic state located in the low-$T$ phase. The physical situation is very different for $d=2$ and $d=3$, which is reflected in the character of the obtained solutions to the RG flow equations. 
We therefore discuss the two cases separately in the two distinct subsections below. 
\subsection{d=2}
In $d=2$ order parameter fluctuations destroy the long-range order in accord with the Mermin-Wagner theorem. A superfluid state may however still exist as a Kosterlitz-Thouless phase\cite{Berezinskii71, Kosterlitz73} 
characterized by algebraically decaying 
order parameter correlations and finite phase stiffness persisting despite vanishing of the expectation value of the order parameter field. In the RG flow the Kosterlitz-Thouless phase is characterized by a line of fixed points parametrized by 
temperature. The superfluid stiffness and the anomalous exponent exhibit jumps (to zero) of universal magnitude as the system crosses the transition temperature $T_{KT}$ and goes into the normal state. In Fig.~1 and Fig.~2 we demonstrate the 
occurrence of the Kosterlitz-Thouless phase by plotting the flow of the anomalous dimension $\eta=-\frac{k}{Z_\pi}\partial_k Z_\pi$ and the superfluid stiffness $\rho_s=2Z_\pi \rho_0$ for low temperatures. 
Within the present truncation the line of fixed-points is recovered 
approximately and is manifested by the occurrence of the quasi-plateaus persisting down to extremely low scales (i.e. large $s$).\cite{Graeter95, Gersdorff01, Jakubczyk14, Jakubczyk16}. 
The deviation from the ideal fixed-point behavior is diminished upon lowering $T$ (see Ref.~[\onlinecite{Jakubczyk17}]). The presence of 
the Laplacian ($Y$) term has, as expected, only minor impact on the flow of the couplings $\rho_0$, $\lambda$, $Z_\pi$ and the universal properties of the Kosterlitz-Thouless phase are the same as obtained in the earlier studies.
In all the 
presented plots the parameters of the initial action are as follows: $Z_\pi=Y=1$, $\rho_0=\frac{1}{2}\alpha_0=\frac{1}{2}$, $\lambda=\frac{1}{4}$. The initial scale $s_{0}$ of the flow is chosen so that the propagators are well gapped by the cutoff and 
the flow becomes completely suppressed at the initial stage. We take $s_{0}=-5$.
\begin{figure}[ht]
\begin{center}
\includegraphics[width=8cm]{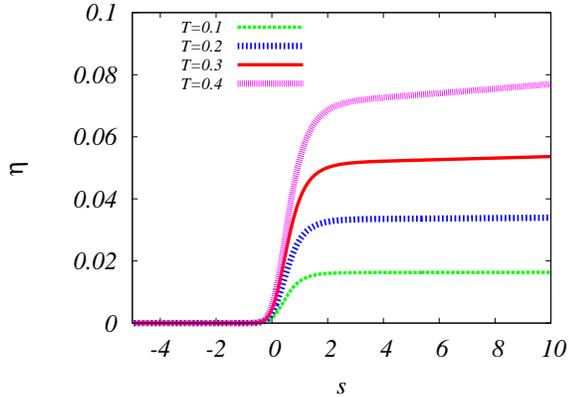}
\caption{The flowing anomalous dimension $\eta$ as a function of the logarithmic scale parameter $s$ for a sequence of temperatures in the low temperature phase in $d=2$. The quasi-plateaus reached in the infrared (large $s$) demonstrate the occurrence of the 
Kosterlitz-Thouless phase. The exponent $\eta$ grows as function of temperature. }
\end{center}
\end{figure}
\begin{figure}[ht]
\begin{center}
\includegraphics[width=8cm]{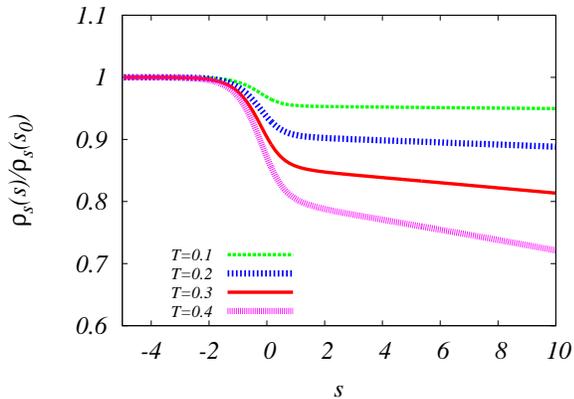}
\caption{The flowing phase stiffness $\rho_s=2 Z_\pi \rho_0$ as a function of the logarithmic scale parameter $s$ for a sequence of temperatures in the low temperature phase in $d=2$. The divergence of $Z_\pi$ (implied by Fig.~1 since $Z_\pi\sim k^{-\eta}$) is 
compensated by the 
vanishing of the order parameter $\rho_0\sim k^{\eta}$ so that $\rho_s = 2 Z_\pi \rho_0$ attains fixed-point like behavior in the infrared (for large $s$) apart from the artificial logarithmic running discussed in Ref.~[\onlinecite{Jakubczyk17}]. }
\end{center}
\end{figure}

In Fig.~3 we illustrate the collapse of the ordering wavevector $\vec{Q_0}$ as anticipated from the structure of the flow equations in Sec.~V and also in full agreement with the reasoning from Sec.~II. Within the range of temperatures used for the plots, the scale of this collapse depends relatively weakly on the temperature $T$ and is comparable to the 
scale where the Kosterlitz-Thouless scaling sets in (see Fig.~1 and Fig.~2). The emergent picture 
also does not strongly depend on the initial choice of the magnitude of $\vec{Q_0}$.
\begin{figure}[ht]
\begin{center}
\includegraphics[width=8cm]{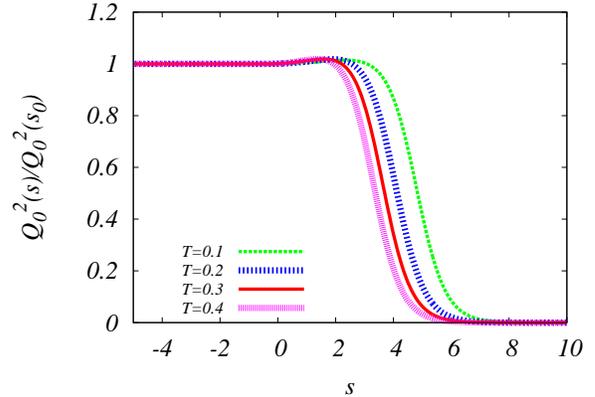}
\caption{The flowing ordering wavevector $\vec{Q_0}^2$ as a function of the logarithmic scale parameter $s$ for a sequence of temperatures in the low temperature phase in $d=2$. The collapse of $\vec{Q_0}^2$ in the infrared (for large $s$) reflects 
the flow properties inferred in 
Sec.~V and implies an instability of the FFLO-type state to order parameter fluctuations.}
\end{center}
\end{figure}
\subsection{d=3}
The physical situation in $d=3$ is quite distinct as compared to $d=2$, since the system admits true long-range order in the low-temperature phase. This is reflected in $Z_\pi$ and the order parameter $\rho_0$ being renormalized 
towards constant, positive values. We demonstrate this in Fig.~4 and Fig.~5. 
\begin{figure}[ht]
\begin{center}
\includegraphics[width=8cm]{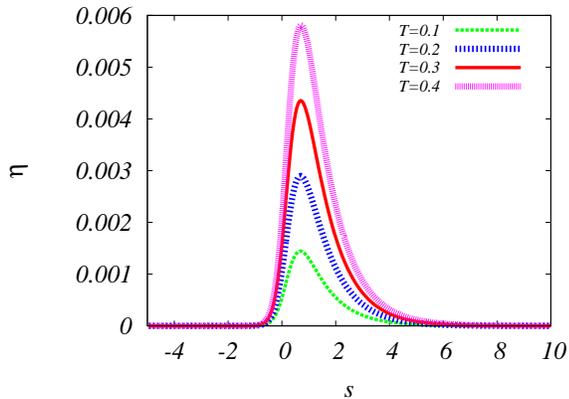}
\caption{The flowing anomalous dimension $\eta$ as a function of the logarithmic scale parameter $s$ for a sequence of temperatures in the low temperature phase in $d=3$. The quantity $Z_\pi$ attains a constant value in the infrared (for large $s$), 
which is signaled by 
vanishing of $\eta$ in contrast to the corresponding behavior in $d=2$. The magnitude of the peak grows upon increasing $T$. At the transition to the normal state, the observed behavior would gradually transform into a a fixed-point reflecting  
 the universal behavior specific to the classical $XY$ model in $d=3$.  }
\end{center}
\end{figure}
\begin{figure}[ht]
\begin{center}
\includegraphics[width=8cm]{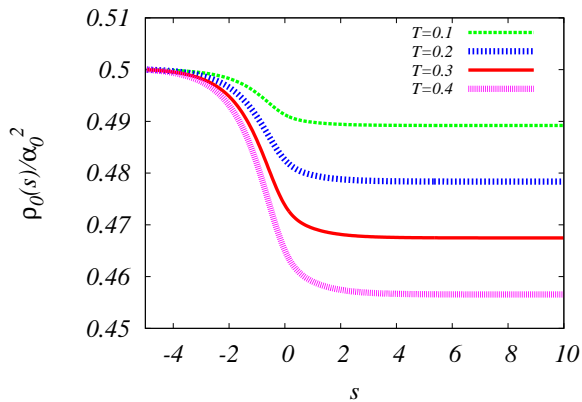}
\caption{The flowing order parameter $\rho_0$ as a function of the logarithmic scale parameter $s$ for a sequence of temperatures in the low temperature phase in $d=3$. The quantity $\rho_0$ becomes renormalized to a reduced, 
but positive value, unlike for  
$d=2$, where it vanishes at any $T>0$. The renormalized value decreases upon rising $T$ and would finally disappear for higher $T$ upon crossing the critical temperature $T_c$. }
\end{center}
\end{figure}
Alike in the previous section, the initial parameters are taken to be $Z_\pi=Y=1$, $\rho_0=\frac{1}{2}\alpha_0=\frac{1}{2}$, $\lambda=\frac{1}{4}$. The presented plots correspond to the system being located deep in the FFLO 
phase at the mean field level. The precise structure of the effective potential (e.g. the presence of competing minima
describing metastable states) is irrelevant for the conclusion. A choice of the initial condition sufficiently close to the normal state (e.g. at higher $T$) 
would lead to a renormalization flow into the normal (i.e. nonsuperfluid) phase with $\rho_0=0$. The effect of renormalizing  $\vec{Q_0}$ to zero is exhibited in Fig.~6.  Due to the divergence of $Y$ in the infrared limit, 
the FFLO-type state cannot be maintained and the wavevector $\vec{Q_0}$ becomes robustly 
renormalized to zero alike in $d=2$. We note however, that the scale where $\vec{Q_0}^2$ vanishes becomes significantly shifted to lower values of $k$ (larger $s$). In a realistic situation this may be of some relevance since a finite $\vec{Q_0}$ may 
remain observed as long as the system size remains sufficiently small. Observe also that the $\vec{Q_0}^2$ collapse scale is by far lower than the scale where the flow of $\rho_0$ freezes in Fig.~2, which also distinguishes $d=3$ from $d=2$. 
\begin{figure}
\begin{center}
\includegraphics[width=8cm]{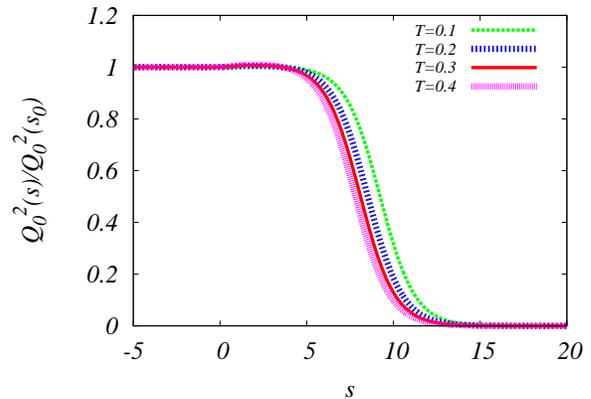}
\caption{The flowing ordering wavevector $\vec{Q_0}^2$ as a function of the logarithmic scale parameter $s$ for a sequence of temperatures in the low temperature phase in $d=3$. The collapse of $\vec{Q_0}^2$ in the infrared (for large $s$) 
reflects the flow properties inferred in 
Sec.~V and implies an instability of the FFLO-type state to order parameter fluctuations. As compared to $d=2$, we observe a significant shift of the collapse scale towards lower values of $k$ (larger $s$). }
\end{center}
\end{figure} 

We would like to emphasize at this point that the mechanism leading to the collapse of $\vec{Q_0}$ is very generic and the calculation may well be relevant also to other instabilities occurring at nonzero $\vec{Q_0}$. An obvious example are spin density waves. One should however realize the limitations of the present approach, which may, perhaps, not be immediately visible. Indeed, as already emphasized, the calculation relies on the derivative expansion, which should, in the limit of $k$ small,  be viewed as a fully trustworthy treatment only for sufficiently small momenta. In consequence, considering $Q_0$ large at the beginning of the flow may lead to a situation, where the theory would be used beyond its applicability range in the limit of $k$ small. The BMW framework, which treats the full momentum dependence of the two-point function at a functional level, is a natural candidate for an extension of the present study, which should be free of this limitation. 


\section{Conclusion}
In this paper we have studied the effect of order parameter fluctuations on the superfluid density wave states known as Fulde-Ferrell-Larkin-Ovchinnikov phases. We set out by reviewing the mean field theory together with the effective theory for the 
Goldstone fluctuations. The structure of this effective action is different for the distinct competing ground states (for example for the FF and LO states). We therefore introduced a simplified model exhibiting weaker fluctuation effects as compared 
to both the FF and LO situations. We argued that in $d=2$ the necessity of $\vec{Q_0}$ being renormalized to zero by fluctuations is implicit in the very structure of the effective action.  We subsequently employed renormalization group theory to 
analyze the system. From the approximate RG flow equations we derived an effect of renormalizing the ordering wavevector to zero for both $d=2$ and $d=3$. The effect 
persists at arbitrarily low (but finite) temperature $T$. The calculation indicates that in these situations the FFLO-type states may be unstable as thermodynamic phases. The mechanism leading to the collapse of the ordering wavevector 
is distinct from those invoked in earlier 
studies and makes no reference to the specific forms of Goldstone spectra present in the FF and LO states, which amplify fluctuation effects.  
Importantly, the scale of vanishing of $\vec{Q_0}$ is much lower in $d=3$ as compared to $d=2$ and it may be that 
the FFLO states are detected in different contexts and identified as true thermodynamic phases due to insufficient system sizes. Remarkably, the experimental evidence for the FFLO type states seems to be available only for charged systems, where the 
Goldstone mode acquires a gap due to the presence of the magnetic field, and to which our study does not apply. There seems to be no indication of the FFLO phases in experiments on cold atoms. Our results seem fully in line with this state of the art.  
We make here no claims concerning $T=0$, where the FFLO phases presumably exist even for neutral systems. 
This is a promising direction for future studies. In realistic situations signatures of such ground states might certainly be detectable also at $T>0$.  


\begin{acknowledgments}
I am grateful to K.~Byczuk, T. Doma\'{n}ski, N. Dupuis, A.~Eberlein, K. Kapcia, W. Metzner, K. Wysoki\'{nski}, H. Yamase, R.~Zeyher for fruitful discussions. I acknowledge support from the Polish National Science Center via grant 
2014/15/B/ST3/02212. I also thank the Erwin-Schr{\"o}dinger Institute in Vienna for hospitality at the initial stage of this work. 
\end{acknowledgments}


\end{document}